\documentclass[utf8]{frontiersFPHY} % for Science, Engineering and Humanities and Social Sciences articles
\usepackage{url,hyperref,lineno,microtype,subcaption}
\usepackage[onehalfspacing]{setspace}
\usepackage{soul}
\usepackage{amsmath}
\usepackage{amssymb}
\usepackage{bm}
\newcommand{\ra}[1]{\textcolor{black}{#1}}
\newcommand{\mh}[1]{\textcolor{black}{#1}}
\newcommand{\magenta}[1]{\textcolor{black}{#1}}

\newcommand{\red}[1]{\textcolor{black}{#1}}

\def\keyFont{\fontsize{8}{11}\helveticabold }
\def\firstAuthorLast{Hirayama {et~al.}} %use et al only if is more than 1 author
\def\Authors{Motoaki Hirayama$^{1,2*}$,  Yusuke Nomura$^{2}$, Ryotaro Arita$^{2,3}$}
\def\Address{$^1$ QPEC, The University of Tokyo, Hongo, Tokyo, 113-8656, Japan \\  
  $^2$  RIKEN Center for Emergent Matter Sciences (CEMS), Wako, Saitama, 351-0198, Japan \\ 
  $^3$  Department of Applied Physics, The University of Tokyo, Hongo, Tokyo, 113-8656, Japan
 }
\def\corrAuthor{Motoaki Hirayama}

\def\corrEmail{hirayama@ap.t.u-tokyo.ac.jp}

\begin{document}
\onecolumn
\firstpage{1}

\title[{\it Ab initio} downfolding based on GW]{{\it Ab initio} downfolding based on the GW approximation for infinite-layer nickelates} 

\author[\firstAuthorLast ]{\Authors} %This field will be automatically populated
\address{} %This field will be automatically populated
\correspondance{} %This field will be automatically populated

\extraAuth{}% If there are more than 1 corresponding author, comment this line and uncomment the next one.
%\extraAuth{corresponding Author2 \\ Laboratory X2, Institute X2, Department X2, Organization X2, Street X2, City X2 , State XX2 (only USA, Canada and Australia), Zip Code2, X2 Country X2, email2@uni2.edu}

\maketitle

\begin{abstract}
%We calculate the effective model for nickel oxides from the band structure in the GW approximation (GWA). We derive a effective 3 band model for the Ni $x^2-y^2$ orbital and the O $2p$ forming the $\sigma$-bond with it. In the GWA, the self-energy correction increases the energy difference between Ni $3d$ and O $2p$, which leads to an decrease in the bandwidth of the antibonding $x^2-y^2$ orbitals and an increase in the bare Coulomb interaction of it. In addition, the screening effect is reduced because the bands in the GWA move away from the Fermi level compared to the LDA. Therefore, the correlation effect of the system is increased compared to that in the LDA, and the system becomes closer to the Mott-Hubbard type model.
\ra{
We derive an effective three-orbital model for the infinite-layer nickelates based on the band structure obtained by the GW approximation (GWA), where we consider the Ni 3$d_{x^2-y^2}$ and O 2$p$ orbitals forming the $\sigma$-bond. In the GWA, the self-energy correction to the local density approximation (LDA) increases the energy difference between Ni $3d_{x^2-y^2}$ and O $2p$, which reduces the bandwidth of the antibonding 3$d_{x^2-y^2}$ orbitals. The isolation of the Ni $3d_{x^2-y^2}$ around the Fermi level 
%enhances the bare Coulomb interaction and 
suppresses the screening effect. 
As a result, the correlation effect becomes more significant than that in the model constructed by the LDA-based downfolding. }
\magenta{
Furthermore, the Mott-Hubbard type character is enhanced in the GWA-based effective model, because the charge-transfer energy increases more rapidly compared to the increase in the interaction parameters. 
}
%suggesting that the system resides deeply in the Mott-Hubbard regime.

\tiny
 \keyFont{ \section{Keywords:} nickelate superconductivity, density functional theory, GW approximation, {\it ab initio} downfolding} %All article types: you may provide up to 8 keywords; at least 5 are mandatory.
\end{abstract}

\section{Introduction}

\ra{
The discovery of nickel superconductors~\cite{Li_2019} has attracted renewed attention to superconductivity in strongly correlated electron systems~\cite{Norman_2020,Pickett_2021,J_Zhang_2021,Botana_2021,nomura2021superconductivity,gu2021superconductivity}. }
\magenta{
So far, superconductivity has been found in film samples of doped infinite-layer nickelates $R$NiO$_2$ ($R$=Nd, Pr, and La)~\cite{Li_2019,Zeng_2020,Q_Gu_2020,Q_Gao_2021,XR_Zhou_2021,Y_Li_2021,Osada_2020,Osada_2020_2,Osada_2021,SW_Zeng_arXiv} and a quintuple-layer nickelate Nd$_6$Ni$_5$O$_{12}$~\cite{Pan_2021}. 
Although the nature of the superconductivity is largely unknown, the pairing mechanism is likely to be unconventional:  
Theoretically, a phonon calculation for NdNiO$_2$ has shown that the electron-phonon coupling is too weak to explain the superconductivity with a transition temperature on the order of 10 K~\cite{Nomura_2019}. 
Experimentally, both $U$- and $V$-shaped spectra have been observed using the scanning tunneling microscopy, depending on the location of the inhomogeneous surface of the doped NdNiO$_2$ film~\cite{Q_Gu_2020}.
Although the origin of the coexistence of the two different signals is controversial~\cite{Adhikary_2020,Z_Wang_2020,Kitamine_2020,X_Wu_arXiv,Choubey_2021}, the presence of the $V$-shape spectrum is consistent with an unconventional $d$-wave pairing. 
In fact, unconventional pairing mechanisms have been discussed since the early stages of the research~\cite{Sakakibara_2020,Hirsch_2019,Wu_2020}.
}

\ra{In contrast with the conventional phonon-mediated superconductivity for which {\it ab initio} calculation based on density functional theory (DFT) plays a crucial role~\cite{RevModPhys.89.015003, FLORESLIVAS20201}, construction of low-energy models with few degrees of freedom is critically important 
for unconventional superconductivity since a detailed analysis of the correlation effects is mandatory. 
In the standard approach to derive a low-energy effective model from first principles, we first calculate the electronic structure with the local density approximation (LDA) or the generalized gradient approximation (GGA) in the framework of DFT. We then construct the maximally localized Wannier function (MLWF)~\cite{Marzari_1997,Souza_2001} for the low-energy states around the Fermi level and derive a tight-binding model. Next, we calculate the effective Coulomb interaction by the constrained random phase approximation (cRPA)~\cite{Aryasetiawan_2004,ImadaMiyake}. The matrix elements of the (partially) screened interaction are calculated for the Wannier basis, from which we estimate the Hubbard $U$ and Hund coupling $J$ in the multi-orbital Hubbard model\cite{Nomura_2019,Sakakibara_2020,Hirayama_2020,Nomura_2020}.
%Another approach evaluating interaction parameters in the effective Hamiltonian is the constrained GW (cGW) method~\cite{hirayama13,Hirayama_2018,Hirayama_2019}, which is free from double counting of the correlation effects on the low-energy degrees of freedom.
The cRPA %and the cGW are 
is formulated in such a way that RPA calculation for the derived low-energy effective model reproduces a one-shot GW ($G_0W_0$) result~\cite{Aryasetiawan_2004,ImadaMiyake,hirayama17}.}

%By using the Green's function of the GW approximation (GWA) beyond the DFT/LDA as $G_0$, we can estimate the screening effects with higher accuracy.
%Actually, the derivation of the effective model from the GW band has been performed for copper oxides~\cite{Hirayama_2018,Hirayama_2019}.
%As in nickel oxides, two types of orbitals, Cu $3d$ and O $2p$ with different localizations, form a band near the Fermi level in copper oxides.
%In the GWA beyond the DFT/LDA, the energy difference between the $d$ and $p$ orbitals
%and the band width are improved
%\mh{is increased and the band width is decreased} 
%by self-energy correction.
%By solving the obtained effective model with a variational Monte Carlo method~\cite{Hirayama_2019},
%the Mott gap and magnetic moment of La$_2$CuO$_4$ successfully reproduce the experimental value and we can characterize superconducting phases in doped systems by \textit{ab initio} manners~\cite{PhysRevB.101.045124}.
%The changes of bands from the DFT/LDA to the GWA is common not only in copper oxides but also in transition metal oxides where $3d$ and $2p$ orbitals with different correlation strengths are hybridized near the Fermi level, and is also expected to produce quantitative changes of the effective model.

\ra{To improve the accuracy of the parameters in the low-energy model, we can replace the Green's function ($G_0$) constructed from the DFT/LDA eigenenergies with the dressed Green's function in the GW approximation (GWA)\footnote{\red{It should be noted that although the cRPA method is free from the double counting problem for the interaction parameters, we have to apply the constrained GW (cGW) method to avoid the double counting in the self-energy~\cite{PhysRevB.87.195144}.}}.
Such a derivation based on the GWA has been recently performed for the celebrated cuprate superconductors~\cite{Hirayama_2018,Hirayama_2019}. While two types of orbitals, i.e., the Cu $3d$ and O $2p$ orbitals, form low-energy bands near the Fermi level, the GW self-energy correction increases the energy difference between the $d$ and $p$ orbitals and reduce the bandwidth of the $d$ band. With these modifications, it has been shown with an extensive variational Monte Carlo \mh{(VMC)} calculation that the experimental values of the Mott gap and magnetic moment of La$_2$CuO$_4$ are successfully reproduced~\cite{Hirayama_2019,PhysRevB.101.045124}. Given that the differences in the band structure between the DFT/LDA and that in the GWA are commonly seen in transition metal oxides where $3d$ and $2p$ orbitals with different correlation strengths coexist near the Fermi level, it would be of great interest to derive an effective low-energy model for infinite-layer nickelates based on the GWA.}

In this study, we perform a first-principles derivation of the effective model for infinite-layer nickelates.
\magenta{
In particular, we mainly focus on the $dpp$ three-orbital models (single-orbital model is discussed in Appendix) because it is interesting to investigate how the GWA modifies the charge-transfer energy and correlation strength compared to the LDA-based downfolding.}\footnote{
\magenta{We note that there are several other effective models for infinite-layer nickelates that have been discussed, including a multi-band model that includes $3d$ orbitals other than the $3d_{x^2-y^2}$ orbital~\cite{Jiang_2020,YH_Zhang_2020,Werner_2020,Petocchi_2020,LH_Hu_2019,Lechermann_2020,Lechermann_2020b,Lechermann_2021,J_Chang_2020,Y_Wang_2020,Z_Liu_2021,B_Kang_arxiv,Choi_2020b}, a model that includes the contribution of rare-earth $4f$ electrons~\cite{P_Jiang_2019,Choi_2020,R_Zhang_2021,Bandyopadhyay_2020}, and a model that includes the self-doping bands~\cite{Hepting_2020,GM_Zhang_2020,Y_Gu_2020}. 
Here, we focus on the debate~\cite{Hepting_2020,Fu_arXiv,Goodge_2021,Jiang_2020,Kitatani_2020,Karp_2020b,Higashi_2021,Karp_2020,ZJ_Lang_2021,worm2021correlations} on the classification of the Mott-Hubbard or charge-transfer regimes in Zaanen-Sawatzky-Allen phase diagram~\cite{Zaanen_1985}. 
}
}
First, we calculate the band structure in the DFT/LDA and estimate the parameter of the effective model using the MLWF and cRPA technique.
Next, we calculate the band structure in the GWA using the Green's function of the LDA.
We derive the effective model from the GW band structure and compare the results with those obtained from the LDA.
\magenta{
We find that the GWA-based effective model is predicted to be more strongly-correlated with enhanced Mott-Hubbard type character.
The model offers an interesting reference to be compared with that of the cuprates with the charge-transfer type character. 
}
%In addition to the simplest single-band model and the $dpp$ three-band model,
%there are several other models of nickel oxides that have been discussed, including a multi-band model that includes $3d$ orbitals other than the $3d_{x^2-y^2}$ orbital, a model that includes the contribution of rare-earth $4f$ electrons, and a model that includes the band originating from the block layer that produces self-doping.
%In this paper, we mainly discuss the three-band model; the single-band model is given in Appendix.
%Just as the effective model from the GW band in the copper oxides has been successful in reproducing experimental values such as the Mott gap, such an effective model would be of great value for nickel oxides.

\section{Method}

In this study, we calculate the parameter of the Hubbard Hamiltonian for the low-energy degree of freedom, 
\begin{eqnarray}
\mathcal{H}^{\text{eff}}= \sum_{ij} \sum_{\ell_1 \ell_2\sigma }&&
t_{\ell_1 \ell_2\sigma}(\bm{R}_i-\bm{R}_j) d_{i\ell_1\sigma} ^{\dagger} d_{j \ell_2\sigma} \nonumber \\
+ \frac{1}{2} \sum_{i_1i_2i_3i_4} %\sum_{klmn \sigma \eta \rho \tau}  
\mh{\sum_{\ell_1\ell_2\ell_3\ell_4 \sigma \eta \rho \tau}}  
&\biggl\{& W^{\text{H}}_{ \ell_1 \ell_2 \ell_3 \ell_4\sigma \eta \rho \tau }(\bm{R}_{i_1},\bm{R}_{i_2},\bm{R}_{i_3},\bm{R}_{i_4}) \nonumber \\
&&d_{i_1 \ell_1\sigma}^{\dagger}d_{i_2 \ell_2\eta} d_{i_3 \ell_3\rho}^{\dagger} d_{i_4 \ell_4\tau}\biggl\} .
\label{Hamiltonian0}
\end{eqnarray}
Here, the hopping term is represented by 
\begin{equation}
t_{ \ell_1 \ell_2\sigma}(\bm{R})= \langle \phi _{ \ell_1\bm{0}}|H|\phi _{ \ell_2\bm{R}} \rangle, 
\label{t}
\end{equation}
where $H$ is the Hamiltonian in the LDA or GWA and $\phi_{\ell\bm{R}}$ is the MLWF of the $\ell$th orbital localized at the unit cell $\bm{R}$. %\ra{It should be noted that we employ the quasi-particle approximation for the self energy in the GWA.}
The interaction term is given by 
\begin{eqnarray}
W_{ \ell_1 \ell_2 \ell_3 \ell_4\sigma \eta \rho \tau }^\text{H}(\bm{R}_{i_1},\bm{R}_{i_2},\bm{R}_{i_3},\bm{R}_{i_4}) 
= \langle \phi _{ \ell_1\bm{R}_{i_1}}\phi _{ \ell_2\bm{R}_{i_2}}|{W^\text{H}}|\phi _{ \ell_3\bm{R}_{i_3}}\phi _{ \ell_4\bm{R}_{i_4}} \rangle ,
\label{WH}
\end{eqnarray}
where $W^\text{H}$ is the effective interaction for the low-energy degree of freedom,
\begin{eqnarray}
W^\text{H} (q,\omega) = \frac{v(q)}{1- P^\text{H}(q,\omega)v(q)}.
\label{WH_def}
\end{eqnarray}

We calculate the effective interaction from the one-shot GWA band.
In the one-shot GWA, we calculate the self-energy from the Green's function $G$ and the fully-screened interaction $W$,
\begin{eqnarray}
\Sigma =GW,
\label{GW}
\end{eqnarray}
where $W$ is calculated from all the polarizations in the RPA $P$ as follows,
\begin{eqnarray}
W (q,\omega) = \frac{v(q)}{1- P(q,\omega)v(q)}.
\label{WH_def}
\end{eqnarray}
The quasiparticle approximation of the Hamiltonian in the GWA is expressed as 
\begin{eqnarray}
H^{\text{GW}}=H^{\text{LDA}}+Z(\epsilon ^{\text{LDA}})(-V^{\text{xc}}+\Sigma(\epsilon ^{\text{LDA}})),
\label{HGW}
\end{eqnarray}
where $H^{\text{LDA}}$ is the Hamiltonian in the LDA, $V^{\text{xc}}$ is the exchange correlation potential in the LDA, and $Z(\epsilon ^{\text{LDA}})$ is the renormalization factor of $\Sigma $ at the eigenenergy $\epsilon ^{\text{LDA}}$: 
\begin{eqnarray}
Z(\epsilon )= \biggl\{ 1-\frac{\partial \text{Re}\Sigma}{\partial \omega }\Big|_{\omega =\epsilon } \biggr\}^{-1}.
\label{Z}
\end{eqnarray}

We calculate the electronic band structure of the YNiO$_2$ using the experimental lattice parameters of LaNiO$_2$, where $a=3.959 \AA$ and $c= 3.375 \AA$~\cite{CRESPIN20051326}. %\magenta{REF}
To exclude the contribution of the $4f$ orbital, here we use Y as the cation.
The computational conditions for the DFT/LDA and GW are as follows.
The calculation is based on the full-potential linear muffin-tin orbital implementation~\cite{methfessel}.
The exchange correlation functional is obtained by the local density approximation of the Ceperley-Alder type~\cite{Ceperley80}.
We neglect the spin-polarization.
The self-consistent LDA calculation is done for the 12 $\times$ 12  $\times$ 12 $k$-mesh.
The muffintin (MT) radii are as follows:
$R^{\text{MT}}_{\text{Y}}=$ 2.9 bohr,
$R^{\text{MT}}_{\text{Ni}}=$ 2.15 bohr,
$R^{\text{MT}}_{\text{O}}=$ 1.5 bohr,
The angular momentum of the atomic orbitals is taken into account up to $l=4$ for all the atoms. 

The cRPA and GW calculations use a mixed basis consisting of products of two atomic orbitals and interstitial plane waves~\cite{schilfgaarde06}.
In the cRPA and GW calculation, the 6 $\times$ 6 $\times$ 6 $k$-mesh is employed for YNiO$_2$.
we interpolate the mesh using the tetrahedron method to treat the screening effect accurately~\cite{fujiwara03,nohara09}.
We disentangle the target band from other bands when the target band crosses another band and construct orthogonalized two separated Hilbert spaces~\cite{miyake09}.
We include bands about from $-25$ eV to $120$ eV for calculation of the screened interaction and the self-energy.

\section{Result}

\begin{figure}[tb]
\vspace{0cm}
\begin{center}
\includegraphics[width=0.30\textwidth]{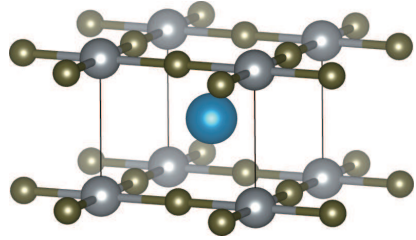}
\caption{
Crystal structure of YNiO$_2$.
}
\label{Fig_structure}
\end{center}
\end{figure}

\begin{figure}[tb]
\vspace{0cm}
\begin{center}
\includegraphics[width=0.95\textwidth]{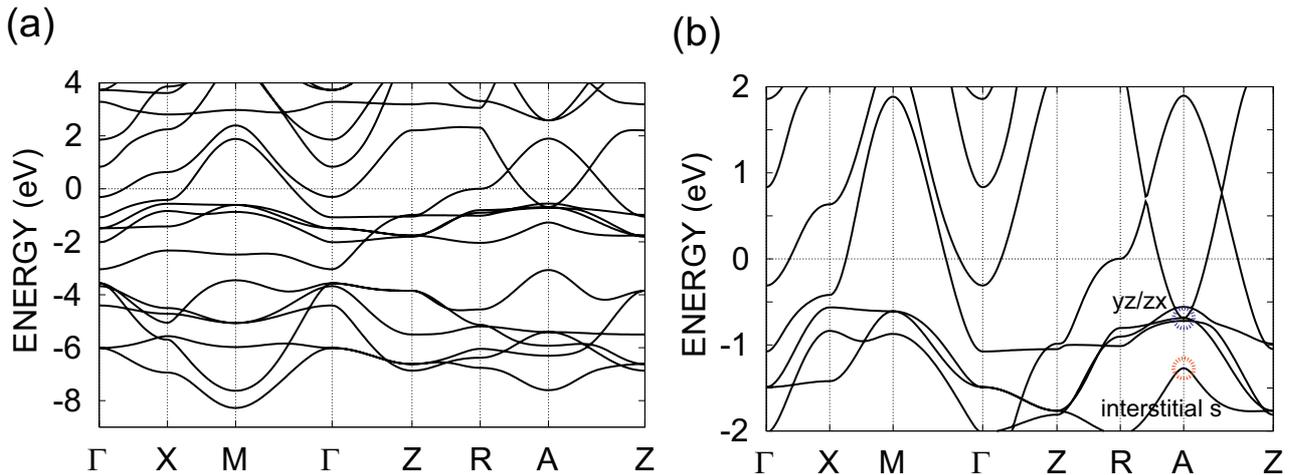}
\caption{\ra{
(a) DFT/LDA band structure for YNiO$_2$ and (b) its magnified figure.
The zero energy corresponds to the Fermi level.
}}
\label{Fig_LDAband_YNiO2}
\end{center}
\end{figure}

\begin{figure}[tb]
\vspace{0cm}
\begin{center}
\includegraphics[width=0.5\textwidth]{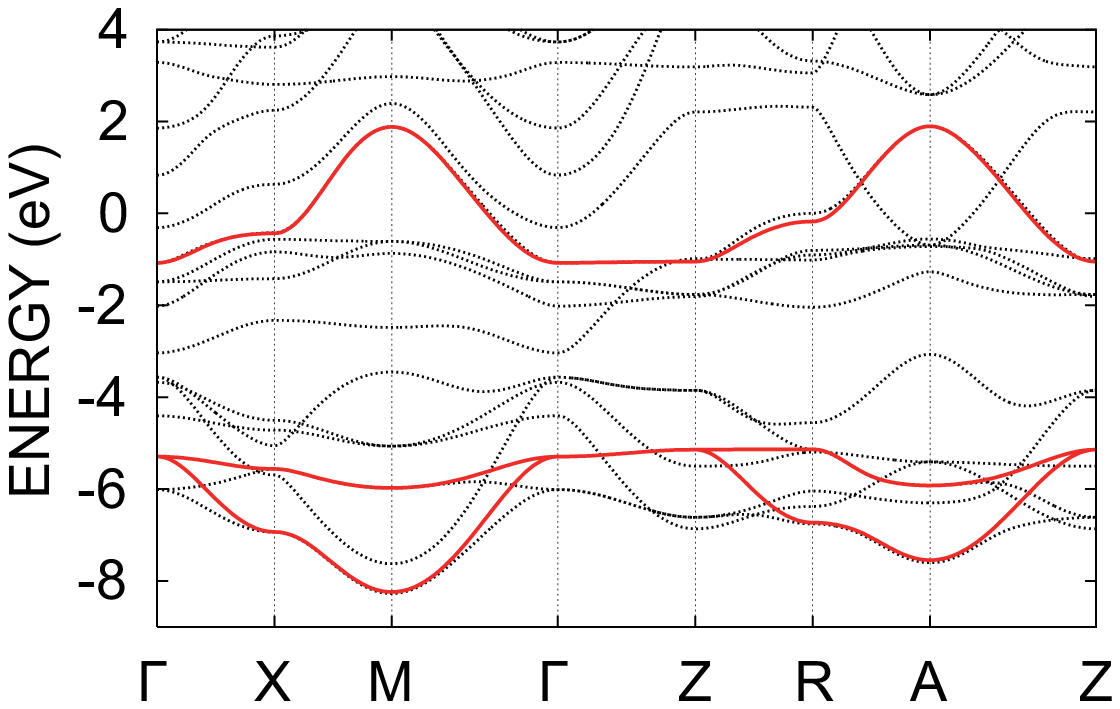}
\caption{\ra{
Electronic band structure of the three-orbital model in the LDA (solid lines).
%Electronic band structure of 3 band model in the LDA originating from the Ni $x^2−y^2$ and the O $2p$ Wannier orbitals.
The zero energy corresponds to the Fermi level.
For comparison, the band structures in the LDA is also given (dotted lines).}
}
\label{Fig_maxloc_LDA}
\end{center}
\end{figure}

\begin{table*}[h]
	\caption{\ra{
		Transfer integrals and effective interactions in the three-orbital Hamiltonian for YNiO$_2$ (in eV).
		Both the one- and two-body part of the Hamiltonian are constructed based on the LDA band structure.
%		while the effective interactions are evaluated by the cRPA calculation for the LDA bands.
		$v$, $U(0)$, \red{$J_v$,} and $J(0)$ represent the bare Coulomb, the static values of the effective Coulomb, \red{bare exchange interactions,} and exchange interactions, respectively (at $\omega=0$). 
		The index 'n' and 'nn' represent the nearest unit cell [1,0,0] and the next-nearest unit cell [1,1,0], respectively. }
	}
	\ 
	\label{para3_LDA} 
	%{\scriptsize %%%%%%%%%%%%%%%%%%%%%%%%%%%%%%%%
	\scalebox{0.8}{
	\begin{tabular}{c|ccc|ccc|ccc|ccc}
		\hline \hline \\ [-8pt]
		$t $(LDA)   &       &  $(0,0,0)$  &       &     & $(1,0,0)$ &    &       & $(1,1,0)$ &      &     &  $(2,0,0)$ &     \\ [+1pt]
		\hline \\ [-8pt]
		&  $x^2-y^2$ &  $p_1$ &  $p_2$ & $x^2-y^2$ &  $p_1$ &  $p_2$ &  $x^2-y^2$ &  $p_1$ &  $p_2$ & $x^2-y^2$ &  $p_1$ &  $p_2$ \\ 
		\hline \\ [-8pt] 
		$x^2-y^2$  & $-$1.377 & $-$1.327 & \phantom{$-$}1.327   &  \phantom{$-$}0.062 & $-$0.018 & $-$0.027 &  \phantom{$-$}0.024 & $-$0.006 &  \phantom{$-$}0.006 &  $-$0.005  & \phantom{$-$}0.001 &  \phantom{$-$}0.000 \\ 
		$p_1$          & $-$1.327&  $-$5.355 & $-$0.671  &  \phantom{$-$}1.327 &  \phantom{$-$}0.043  & \phantom{$-$}0.671  & $-$0.027 &  \phantom{$-$}0.037 &  \phantom{$-$}0.002 & \phantom{$-$}0.018 & $-$0.006  & \phantom{$-$}0.002 \\
		$p_2$          &  \phantom{$-$}1.327 & $-$0.671 & $-$5.355  &   $-$0.027 & $-$0.002 & $-$0.043 &  \phantom{$-$}0.027 &  \phantom{$-$}0.002  & \phantom{$-$}0.037 & \phantom{$-$}0.000 &  \phantom{$-$}0.000 &  \phantom{$-$}0.000    \\
		\hline \hline \\ [-8pt]  
		&       &  $v$  &       &     & $U(0)$ &    &       & $J_{v}$ &      &     &  $J(0)$ &     \\ [+1pt]
		\hline \\ [-8pt]
		&  $x^2-y^2$ &  $p_1$ &  $p_2$ & $x^2-y^2$ &  $p_1$ &  $p_2$ &  $x^2-y^2$ &  $p_1$ &  $p_2$ & $x^2-y^2$ &  $p_1$ &  $p_2$ \\ 
		\hline \\ [-8pt] 
		$x^2-y^2$ &  26.406 &  7.886  & 7.886   & 4.599 &  0.763  & 0.763 &            &  0.116 &   0.116  &           &    0.066  &  0.066   \\ 
		$p_1$         &  7.886 & 17.231 &  5.278  &0.763  & 4.127 &  0.499  &  0.116 &            &  0.040  &  0.066   &            &  0.019 \\
		$p_2$        &   7.886  & 5.278  & 17.231  & 0.763 &  0.499   & 4.127 & 0.116   & 0.040  &            &   0.066   & 0.019  &         \\
		\hline \hline \\ [-8pt]  
		&       &  $v_{\text{n}}$ &    &     & $V_{\text{n}}(0)$ &    &       & $v_{\text{nn}}$  &      &     &  $V_{\text{nn}}(0)$ &     \\ [+1pt]
		\hline \\ [-8pt] 
		&  $x^2-y^2$ &  $p_1$ &  $p_2$ & $x^2-y^2$ &  $p_1$ &  $p_2$ &  $x^2-y^2$ &  $p_1$ &  $p_2$ & $x^2-y^2$ &  $p_1$ &  $p_2$ \\ 
		\hline \\ [-8pt] 
		$x^2-y^2$ & 3.730  & 7.886 &  3.286 & 0.157 &  0.763 &  0.124 &  2.644 &  3.286  & 3.286  &0.061 &  0.124 &  0.124    \\
		$p_1$     & 2.530 &  3.841 &  2.379 & 0.080 &  0.250 &  0.059 & 2.124  & 2.643&   2.379  & 0.035 &  0.086   &0.059  \\
		$p_2$        & 3.286  & 5.278 &  3.566 &  0.124 &  0.499 &  0.155 & 2.124  & 2.379 &  2.643  &  0.035  & 0.059   &0.086  \\
		\hline
		\hline 
	\end{tabular} 
	}
	%} %%%%%%%%%%%%%%%%%%
\end{table*}

Figure~\ref{Fig_structure} shows the crystal structure of the infinite-layer nickelates.
The block layer is a single lanthanide cation and has 
\magenta{large interstitial regions surrounded by cations.}
%the large region of the void.
This is one of the reasons for the formation of electron pockets originating from the block layer, as described below.

Figure~\ref{Fig_LDAband_YNiO2} shows the band structure of YNiO$_2$ in the LDA.
The band structure of YNiO$_2$ is \ra{very similar to}
%qualitatively the same as 
that of NdNiO$_2$ \ra{if we eliminate the Nd $4f$ bands}. 
%in the absence of the $4f$ orbitals.
The $3d_{x^2-y^2}$ antibonding state mainly forms the Fermi surface, which is \ra{a feature commonly seen} in the cuprate superconductors.
Reflecting the \magenta{square planar} crystal field of oxygen around the nickel site, the other $d$ bands are \ra{almost fully occupied}. %located in the occupied band.
However, differently from the cuprates, the infinite-layer nickelates have additional small electron pockets around the $\Gamma$ and A points.
These electron pockets originate from the $d$-orbital and the interstitial state in the block layer, respectively.
The energy difference between the $3d$ bands of Ni$^{1+}$ and the $2p$ bands of O$^{2-}$ is larger than that between Cu$^{2+}$ and O$^{2-}$ in copper oxides, and they are energetically separated near $-3$ eV.

The interstitial state is located at $-1.4$ eV at the $A$ point, and has a band inversion between $yz/zx$ orbitals around the $A$ point.
Because of the inversion between bands with different numbers of degeneracies, the bands of the interstitial $s$ and the $yz/zx$ are continuously connected from the conduction band to the valence band.
Since this band inversion is buried in the metallic band, it will be difficult to observe the surface state associated with the band inversion.

In this paper, we derive a three-orbital effective model consisting of the Ni $3d_{x^2-y^2}$ orbital and two O $2p$ orbitals forming a $\sigma$-bonding. % \ra{and anti-bonding bands?}. 
%with the Ni $3d_{x^2-y^2}$ orbital.
%We construct a 3-band model from 11 bands originating from the Ni $3d$ and O $2p$ orbitals near the Fermi level using the maximally localized Wannier function method (Fig.~\ref{Fig_maxloc_LDA}).
\ra{We first construct the maximally localized Wannier functions~\cite{Marzari_1997,Souza_2001} for these orbitals and evaluate the parameters in the tight-binding model (see Table~\ref{para3_LDA}). }
%Table~\ref{para3_LDA} shows the corresponding hopping parameters of the 3 band model.
%The obtained values are in good agreement with previous studies of NdNiO$_2$ and LaNiO$_2$.
\mh{The obtained model has a larger energy difference between the $3d_{x^2-y^2}$ and $2p$ orbitals than that of the cuprate, and is closer to the Mott-Hubbard type.}

We then calculate the effective interaction for the three-orbital model by the cRPA method.
The obtained effective interactions are summarized in Table~\ref{para3_LDA}.
%As with hopping, the results are in good agreement with previous studies.
\mh{The bare Coulomb interaction $v$ is slightly smaller than that of the copper oxides (Ni $3d_{x^2-y^2}$:$\sim$26 eV, Cu $d_{x^2-y^2}$:$\sim$29 eV in Refs.~\cite{Hirayama_2018,Hirayama_2019}),
and the dielectric constant $U/v$ is smaller than that of the copper oxides partially due to the metallic screening from the block layer.}

\begin{figure}[tb]
\vspace{0cm}
\begin{center}
\includegraphics[width=0.95\textwidth]{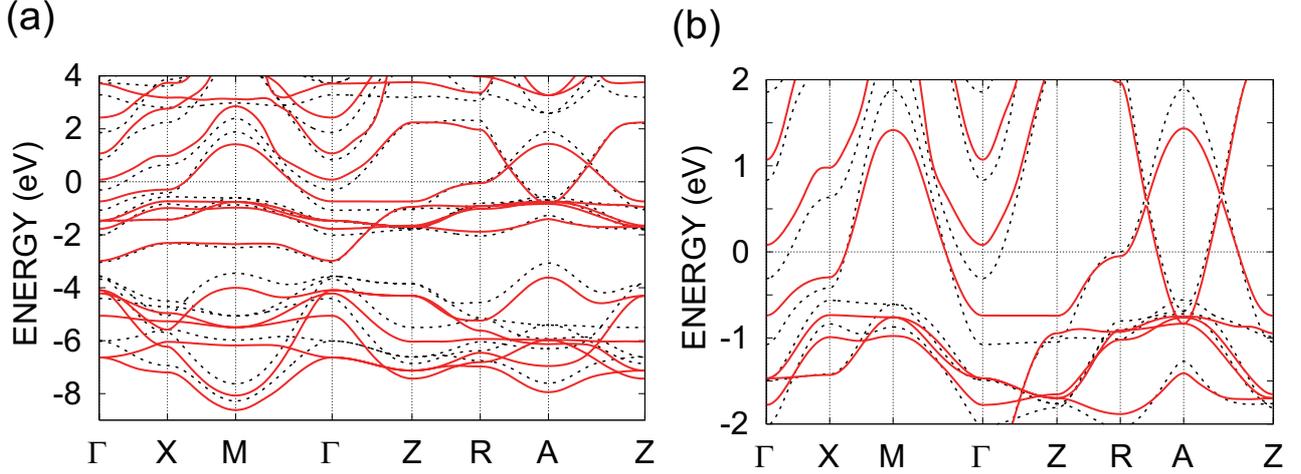}
\caption{\ra{
(a) GW band structure for YNiO$_2$ and (b) its magnified figure (solid lines).
For comparison, the band structures in the LDA
is also given (dotted lines).
The zero energy corresponds to the Fermi level.
}}
\label{Fig_GWband_YNiO2}
\end{center}
\end{figure}

\begin{figure}[tb]
\vspace{0cm}
\begin{center}
\includegraphics[width=0.5\textwidth]{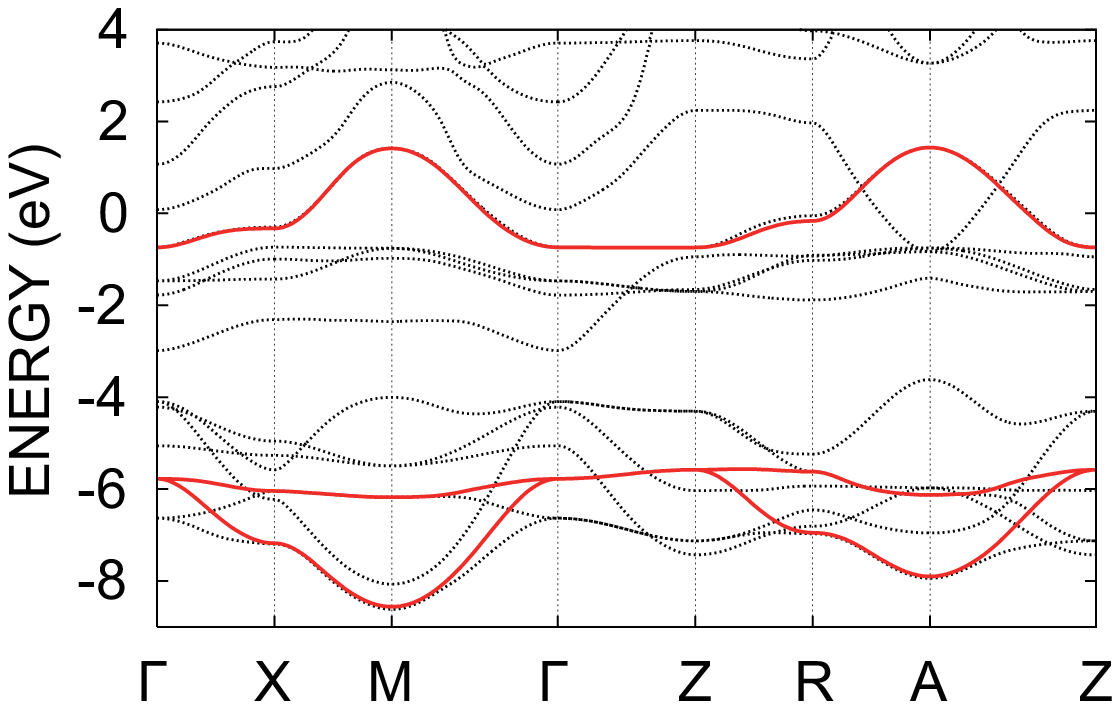}
\caption{\ra{
Electronic band structure of the three-orbital model in the GWA (solid lines).
%Electronic band structure of $dpp$ model in the GW originating from the Ni $x^2−y^2$ and the O $2p$ Wannier orbitals.
The zero energy corresponds to the Fermi level.
For comparison, the band structures in the GWA is also given (dotted lines).
}}
\label{Fig_maxloc_GW}
\end{center}
\end{figure}

\begin{table*}[h] 
	\caption{\ra{
		Transfer integrals and effective interactions in the three-band Hamiltonian for YNiO$_2$ (in eV).
		The one-body part is obtained from the GW band structure,
%		while the effective interaction is the result of the cRPA calculation for the LDA bands.
		\mh{
		and the effective interaction is the result of the cRPA calculation for the GW bands.
		}	
		$v$, $U(0)$, \red{$J_v$,} and $J(0)$ represent the bare Coulomb, the static values of the effective Coulomb, \red{bare exchange interactions,} and exchange interactions,
		respectively (at $\omega=0$). 
		The index 'n' and 'nn' represent the nearest unit cell [1,0,0] and the next-nearest unit cell [1,1,0] respectively. }
	}
	\ 
	\label{para3_GW} 
	%{\scriptsize %%%%%%%%%%%%%%%%%%%%%%%%%%%%%%%%
    \scalebox{0.8}{
	\begin{tabular}{c|ccc|ccc|ccc|ccc} 
		\hline \hline \\ [-8pt]
		$t $(GW)   &       &  $(0,0,0)$  &       &     & $(1,0,0)$ &    &       & $(1,1,0)$ &      &     &  $(2,0,0)$ &     \\ [+1pt]
		\hline \\ [-8pt]
		&  $x^2-y^2$ &  $p_1$ &  $p_2$ & $x^2-y^2$ &  $p_1$ &  $p_2$ &  $x^2-y^2$ &  $p_1$ &  $p_2$ & $x^2-y^2$ &  $p_1$ &  $p_2$ \\ 
		\hline \\ [-8pt] 
		$x^2-y^2$  & $-$1.204 & $-$1.288 &  \phantom{$-$}1.288   &  \phantom{$-$}0.094  & $-$0.025 & $-$0.021 &  \phantom{$-$}0.015 & $-$0.005 &  \phantom{$-$}0.005 &   $-$0.002  & \phantom{$-$}0.001  & \phantom{$-$}0.001 \\ 
		$p_1$          & $-$1.288 & $-$5.802 & $-$0.640  &  \phantom{$-$}1.288  & \phantom{$-$}0.037  & \phantom{$-$}0.640  & $-$0.021  & \phantom{$-$}0.031 &  \phantom{$-$}0.007 & \phantom{$-$}0.025 & $-$0.004  & \phantom{$-$}0.007 \\
		$p_2$          & \phantom{$-$}1.288 &  $-$0.640 & $-$5.802  &    $-$0.021  &$-$0.007 & $-$0.022 &  \phantom{$-$}0.021 &  \phantom{$-$}0.007 &  \phantom{$-$}0.031 & \phantom{$-$}0.001  & \phantom{$-$}0.000  &$-$0.003    \\
		\hline \hline \\ [-8pt]  
		&       &  $v$  &       &     & $U(0)$ &    &       & $J_{v}$ &      &     &  $J(0)$ &     \\ [+1pt]
		\hline \\ [-8pt]
		&  $x^2-y^2$ &  $p_1$ &  $p_2$ & $x^2-y^2$ &  $p_1$ &  $p_2$ &  $x^2-y^2$ &  $p_1$ &  $p_2$ & $x^2-y^2$ &  $p_1$ &  $p_2$ \\ 
		\hline \\ [-8pt] 
		$x^2-y^2$ & 26.596  & 7.901 &  7.901   & 5.019  & 0.932  & 0.932 &            &  0.114  & 0.114  &           &   0.066 &  0.066  \\ 
		$p_1$         &  7.901 & 17.383 &  5.280  & 0.932   &4.510 &  0.624  &  0.114  &            &  0.038  & 0.066  &            &    0.019 \\
		$p_2$        &   7.901 &  5.280 & 17.382  &0.932  & 0.624 &  4.510 &  0.114   & 0.038  &            & 0.066  &   0.019  &         \\
		\hline \hline \\ [-8pt]  
		&       &  $v_{\text{n}}$ &    &     & $V_{\text{n}}(0)$ &    &       & $v_{\text{nn}}$  &      &     &  $V_{\text{nn}}(0)$ &     \\ [+1pt]
		\hline \\ [-8pt] 
		&  $x^2-y^2$ &  $p_1$ &  $p_2$ & $x^2-y^2$ &  $p_1$ &  $p_2$ &  $x^2-y^2$ &  $p_1$ &  $p_2$ & $x^2-y^2$ &  $p_1$ &  $p_2$ \\ 
		\hline \\ [-8pt] 
		$x^2-y^2$ & 3.727 &  7.901 &  3.285 & 0.223 &  0.932  & 0.181 & 2.643 &  3.285 &  3.285  & 0.094 &  0.181 &  0.181    \\
		$p_1$         &  2.528  & 3.840 &  2.379 & 0.116 &  0.332 &  0.094 & 2.123  & 2.641 &  2.379  & 0.057 &  0.130 &  0.094  \\
		$p_2$        &  3.285 &  5.280 &  3.567 & 0.181 &  0.624 &  0.230 & 2.123  & 2.379  & 2.641   &  0.057  & 0.094 &  0.13   \\
		\hline
		\hline 
	\end{tabular} 
	}
	%} %%%%%%%%%%%%%%%%%%
\end{table*}

We next show the band structure in the GWA in Fig.~\ref{Fig_GWband_YNiO2}.
In the GWA, the energy difference between the strongly correlated Ni $3d$ orbitals and the weakly correlated O $2p$ orbitals is enhanced\mh{~\cite{Hirayama_2020,PhysRevB.101.161102}}.
%Therefore, the overall bandwidth of the 11 orbitals near the Fermi level is enhanced.
Thereby, the energy gap between the $d$- and $p$-bands around $-3$ eV is increased.
On the other hand, the bandwidth of the antibonding orbitals of the $3d_{x^2-y^2}$ orbital decreases.
The contribution of the O $2p$ orbitals to the antibonding orbitals decreases due to the increase in the energy difference between the $d$- and $p$-orbitals.
The bandwidth of the strongly correlated orbitals in the GWA is also reduced compared to that in the LDA due to the effect of the frequency dependence of the self-energy.
The bandwidth of the O $2p$ orbitals remains approximately the same as that in the LDA.

In the GWA, the position of the valence band is lifted up from that in the LDA.
In particular, the electron pocket originating from the $d$ orbital in the block layer near the $\Gamma$ point disappears.
On the other hand, the bottom of the band originating from the interstitial state still creates the electron pocket around the $A$ point even in the GWA.

%We calculate the effective interaction of the three-orbital model in the GWA.
We summarize the hopping parameters in Table~\ref{para3_GW}.
The difference in the on-site potential between the $3d_{x^2-y^2}$ and $2p$ orbitals increase from 3.98 eV to 4.60 eV.
The nearest-neighbor hopping between the $3d_{x^2-y^2}$ and $2p$ orbitals is almost the same ($\sim -1.3$ eV), but slightly reduced due to the renormalization factor in the GWA.
The increase of the onsite potential deference between the atomic $3d_{x^2-y^2}$ and $2p$ orbitals results in an decrease of the oxygen contribution to the antibonding $3d_{x^2-y^2}$ orbitals and decrease of the hopping between the antibonding $3d_{x^2-y^2}$ orbitals.

%The screening effect of the system is greatly reduced compared to that in the LDA due to the disappearance of the metallic screening from the electron pocket at the $\Gamma$ point and the decrease in the energy position of the O $2p$ band.
The screening effect of the system is reduced compared to that in the LDA 
mainly due to \magenta{the increase of the charge-transfer energy,} 
%decrease in the energy position of the O $2p$ band, 
which \red{increases} the bare Coulomb interaction of the $3d_{x^2-y^2}$ band and reduces the screening effect from the $2p$ bands.
\ra{The bands \mh{originating from the block layer as well as the O $2p$ orbitals} in the GWA move away from the Fermi level compared to the LDA, which makes the screening effect weaker.}
The disappearance of the metallic screening from the electron pocket at the $\Gamma$ point also partially contribute to the reduction of the correlation.
Therefore, the value of the effective interaction is increased from that in the LDA.
For example, \ra{while the on-site interaction is 4.6 eV for the $3d_{x^2-y^2}$ orbital and 4.1 eV for the $2p$ orbital in the LDA-based cRPA calculation, the GWA-based cRPA gives 5.0 eV for the $3d_{x^2-y^2}$ orbital and 4.5 eV for the $2p$ orbital.}
The nearest-neighbor interactions also increase from 0.16 eV to 0.22 eV for the $3d_{x^2-y^2}$ orbital.
Note that the metallic screening from the electron pocket near the A point still remains even in the GWA.\footnote{
\magenta{
We note that there is a proposal that the electron pocket at the $A$ point can be eliminated by designing a different type of the block layer~\cite{Hirayama_2020}.
}
}

\section{Conclusion}
%We derive the effective model from the band structure in the GWA. In the GWA, the O $2p$ bands move away from the Fermi level and the bandwidth of the $x^2-y^2$ band decreases. In GWA, the bands move away from the Fermi level compared to the LDA and the screening effect decreases. Therefore, the effective 3 band model has smaller hopping, larger interactions, and thus stronger correlations than that obtained from the LDA.
\ra{We derived a three-orbital low-energy model for the infinite-layer nickelates based on the GWA. In the GWA, the O $2p$ bands locate deeper below the Fermi level, and the bandwidth of the Ni $3d_{x^2-y^2}$ band is narrower than that in the LDA calculation.
Due to the isolation of the low-energy Ni $3d_{x^2-y^2}$ band, the screening effect becomes less effective, leading to larger interaction parameters in the Hamiltonian. Thus the GW-based {\it ab initio} downfolding gives a more correlated model than the LDA-based downfolding.}

\section*{Appendix}
\ra{
For reference, we summarize the parameters in the single-orbital model in Tables~\ref{para1_LDA} and ~\ref{para1_GW}.
%Note that the strength of the interaction in the GWA should be payed to attention (see Ref.~\cite{Hirayama_2019}).
Special attention should be paid to the strength of the interaction in the GWA-based effective single-orbital model (see Ref.~\cite{Hirayama_2019}).}
\mh{In the copper oxides, the correlation effect beyond the RPA between the $d$ and $p$ orbitals in the three-orbital model is not small.
Therefore, in order to calculate the single-orbital model accurately,
it is necessary to treat the screening effect originating from the bonding and nonbonding bands beyond the RPA.
To do so, we need to solve the three-orbital model once with a low-energy solver such as the VMC and estimate the energy corrections between the $d$ and $p$ orbitals beyond the GWA.
By combining such a correction with the GW self-energy correction,
we can calculate the band structure beyond the GWA,
and can estimate a single-orbital model with high accuracy (See Ref.~\cite{Hirayama_2019} for details of the method).
Because the nickelates have a qualitatively similar band structure to the cuprates, the reliability of the GWA-based single-orbital model for the nickelates also needs to be carefully examined: in particular, the correlation strength $|U/t|$ might be overestimated.
}
%\mh{In the copper oxides, considering} the correlation effect from bonding and non-bonding states of the O $2p$ orbitals beyond the GWA and the feedback effects from the high energy degrees of freedom, the correlation of the one-band model decreases significantly from that simply estimated from the GW band.\ra{Can we say this from Tables 3 and 4?}
%\mh{
%By starting from a 3 band model and calculating such a correction effect, we can calculate a highly accurate 1 band model~\cite{Hirayama_2019}.
%Nickel oxides have a qualitatively similar band structure to copper oxides, and the correlation effect in the 1 band model might be overestimated.
%}

\begin{table*}[ptb] 
	\caption{
		Transfer integral and effective interaction in the one-band Hamiltonian for YNiO$_2$ (in eV).
		\ra{Both the one-body and two-body parts in the Hamiltonian are derived based on the LDA band structure.}
%        The one-body part is obtained from the fitting of the LDA band structure,
%		while the effective interaction is the result of the cRPA form the LDA bands.
		$v$ and $U(0)$ represent the bare Coulomb interaction and the static value of the effective Coulomb interaction, respectively (at $\omega=0$). 
		The index 'n' and 'nn' represent the nearest unit cell [1,0,0] and the next-nearest unit cell [1,1,0], respectively. 
	}
\ 
\label{para1_LDA} 
%{\scriptsize %%%%%%%%%%%%%%%%%%%%%%%%%%%%%%%%
\begin{tabular}{c|c|c|c|c|c|c} 
	\hline \hline \\ [-8pt]
	$t $(LDA)   &      $(0,0,0)$  &   $(1,0,0)$ &     $(1,1,0)$ &      $(2,0,0)$  & $U/v$ & $|U/t|$  \\ [+1pt]
	\hline \\ [-8pt] 
	$x^2-y^2$   & 0.211 & $-$0.357 & 0.093 & $-$0.046 & 0.149 & 8.15  \\
	\hline \hline \\ [-8pt]  
	&      $v$  &     $U(0)$ &     $v_{\text{n}}$ &    $V_{\text{n}}(0)$ &   $v_{\text{nn}}$  &     $V_{\text{nn}}(0)$    \\ [+1pt]
	\hline \\ [-8pt]
	$x^2-y^2$   &  19.578  &2.910 &  3.981  & 0.229 & 2.685  & 0.091 \\
	\hline
	\hline 
\end{tabular} 
%} %%%%%%%%%%%%%%%%%%
\end{table*}

\begin{table*}[ptb] 
	\caption{
		Transfer integral and effective interaction in the one-band Hamiltonian for YNiO$_2$ (in eV).
		\ra{The one-body part is derived based on the GW band structure,
        %The one-body part is obtained from the fitting of the LDA band structure,
		%while the effective interaction is the result of the cRPA form the LDA bands.
%		while the effective interaction is calculated by the cRPA calculation for the LDA band structure.
		\mh{
		and the effective interaction is the result of the cRPA calculation for the GW bands.
		}
}
		%$v$, $U(0)$ and $J(0)$ represent the bare Coulomb, the static values of the effective Coulomb, and exchange interactions, respectively (at $\omega=0$). 
        $v$ and $U(0)$ represent the bare Coulomb interaction and the static value of the effective Coulomb interaction, respectively (at $\omega=0$).
		The index 'n' and 'nn' represent the nearest unit cell [1,0,0] and the next-nearest unit cell [1,1,0] respectively. 
	}
\ 
\label{para1_GW} 
%{\scriptsize %%%%%%%%%%%%%%%%%%%%%%%%%%%%%%%%
\begin{tabular}{c|c|c|c|c|c|c} 
	\hline \hline \\ [-8pt]
	$t $(GW)   &      $(0,0,0)$  &   $(1,0,0)$ &     $(1,1,0)$ &      $(2,0,0)$  & $U/v$ & $|U/t|$  \\ [+1pt]
	\hline \\ [-8pt] 
	$x^2-y^2$   & 0.172 &$-$0.271 &  0.075 & $-$0.033 & 0.167 & 12.94  \\
	\hline \hline \\ [-8pt]  
	&      $v$  &     $U(0)$ &     $v_{\text{n}}$ &    $V_{\text{n}}(0)$ &   $v_{\text{nn}}$  &     $V_{\text{nn}}(0)$    \\ [+1pt]
	\hline \\ [-8pt]
	$x^2-y^2$   &  20.948  &3.508 &  3.957  & 0.300 & 2.677  & 0.131 \\
	\hline
	\hline 
\end{tabular} 
%} %%%%%%%%%%%%%%%%%%
\end{table*}

\section*{Conflict of Interest Statement}
%All financial, commercial or other relationships that might be perceived by the academic community as representing a potential conflict of interest must be disclosed. If no such relationship exists, authors will be asked to confirm the following statement: 
The authors declare that the research was conducted in the absence of any commercial or financial relationships that could be construed as a potential conflict of interest.

\section*{Author Contributions}
\magenta{MH conducted calculations.}
All authors contributed to writing the article.

\section*{Funding}
\ra{We acknowledge funding through Grant-in-Aids for Scientific Research (JSPS KAKENHI) \magenta{[Grant No. 20K14423 (YN), 21H01041 (YN), and 19H05825 (RA)]} and ``Program for Promoting Researches on the Supercomputer Fugaku'' \magenta{(Basic Science for Emergence and Functionality in Quantum Matter ---Innovative Strongly-Correlated Electron Science by Integration of ``Fugaku'' and Frontier Experiments---) (Grant No. JPMXP1020200104)} from MEXT.}

\section*{Acknowledgments}
We thank Terumasa Tadano and Motoharu Kitatani for valuable discussions.

%\section*{Data Availability Statement}
%The present paper reviews previous work.  Datasets are accessible through the original publications.
% Please see the availability of data guidelines for more information, at https://www.frontiersin.org/about/author-guidelines#AvailabilityofData

%\bibliographystyle{frontiersinSCNS_ENG_HUMS} % for Science, Engineering and Humanities and Social Sciences articles, for Humanities and Social Sciences articles please include page numbers in the in-text citations
\bibliographystyle{frontiersinHLTH&FPHY} % for Health, Physics and Mathematics articles
\bibliography{main}

%%% Make sure to upload the bib file along with the tex file and PDF
%%% Please see the test.bib file for some examples of references

%%% Please be aware that for original research articles we only permit a combined number of 15 figures and tables, one figure with multiple subfigures will count as only one figure.
%%% Use this if adding the figures directly in the mansucript, if so, please remember to also upload the files when submitting your article
%%% There is no need for adding the file termination, as long as you indicate where the file is saved. In the examples below the files (logo1.eps and logos.eps) are in the Frontiers LaTeX folder
%%% If using *.tif files convert them to .jpg or .png
%%%  NB logo1.eps is required in the path in order to correctly compile front page header %%%

%%% If you are submitting a figure with subfigures please combine these into one image file with part labels integrated.
%%% If you don't add the figures in the LaTeX files, please upload them when submitting the article.
%%% Frontiers will add the figures at the end of the provisional pdf automatically
%%% The use of LaTeX coding to draw Diagrams/Figures/Structures should be avoided. They should be external callouts including graphics.

%\bibdata{main}% Produces the bibliography via BibTeX.

\end{document}